# Purcell enhancement and spin spectroscopy of silicon vacancy centers in silicon carbide using an ultra-small mode-volume plasmonic cavity

Jae-Pil So[1], Jialun Luo[2], Jaehong Choi[1], Brendan McCullian[1], and Gregory D. Fuchs[1,3,*]

[1]School of Applied and Engineering Physics, Cornell University, Ithaca, NY 14850, USA

[2]Department of Physics, Cornell University, Ithaca, NY 14850, USA

[3]Kavli Institute at Cornell for Nanoscale Science, Ithaca, NY 14850, USA

[*]Corresponding authors. E-mail: gdf9@cornell.edu

**Abstract**

**Silicon vacancy ($V_{Si}$) centers in 4H-silicon carbide have emerged as a strong candidate for quantum networking applications due to their robust electronic and optical properties including a long spin coherence lifetime and bright, stable emission. Here, we report the integration of $V_{Si}$ centers with a plasmonic nanocavity to Purcell enhance the emission, which is critical for scalable quantum networking. Employing a simple fabrication process, we demonstrate plasmonic cavities that support a nanoscale mode volume and exhibit an increase in the spontaneous emission rate with a measured Purcell factor of up to 48. In addition to investigating the optical resonance modes, we demonstrate that an improvement in the optical stability of the spin-preserving resonant optical transitions relative to the radiation-limited value. The results highlight the potential of nanophotonic structures for** **advancing quantum networking technologies and emphasizes the importance of optimizing emitter-cavity interactions for efficient quantum photonic applications.**

**Introduction**

Solid-state quantum defects with spin and coherent light emission have been extensively investigated as potential building blocks for applications including nanoscale quantum sensing[1-4], quantum computing[5,6] and quantum communications.[7-10] Among these, silicon vacancy ($V_{Si}$) centers in 4H-silicon carbide (4H-SiC) have emerged as promising due to their electronic and optical properties including a long spin coherence lifetime, a strong zero photon line (ZPL) with a large Debye-Waller factor in the range of 6-9%, and a remarkable spectral stability that motivates the integration of these defects with photonic structures.[11-16] These characteristics make $V_{Si}$ centers excellent candidates for applications in quantum networking. Moreover, $V_{Si}$ centers possess optically accessible spin states, which is critical for readout and photon-mediated entanglement of a long-lived spin.[17-20] Combining these characteristics with the excellent optical properties of 4H-SiC including a strong second-order optical nonlinearity,[21] a high index of refraction, a wide bandgap,[22] and the feasibility of silicon compatible fabrication technologies,[23,24] $V_{Si}$ centers emerge as a promising solid-state defect to realize integrated quantum network nodes. Like many quantum photonic systems, enhancing the interaction between $V_{Si}$ centers and photons is crucial for developing scalable and efficient photonic quantum computing and error-corrected quantum communications.[25]

Efforts to integrate nanophotonic cavities with spin defects have been ongoing, especially focused on defect centers in diamond including the nitrogen-vacancy (NV) center and the silicon-vacancy (SiV) center.[26-28] Recently, host materials that have fabrication techniques that are compatible with silicon photonics or III-V material-based photonic structures have been investigated with the aim of creating high-quality interactions between an emitter and a cavity. These studies have focused on integrating $V_{Si}$ with external photonic structures based on optical confinement in the dielectric host material,[18,23,29-31] but these often result in large volumes or require complex freestanding structures that are achieved through selective chemical etching and wafer-bonding. The complication of such approaches is disadvantageous, highlighting the need for simpler integration methods. One promising approach involves the use of surface plasmon resonance (SPR) in metallic cavities, which can achieve extremely small mode volumes due to strong light-matter interactions and high-field enhancements.[32] This method has been extensively reported for miniaturizing lasers,[33,34] sensors,[35] and high-speed devices,[36] and notably, for enhancing quantum emitters through the Purcell effect.[32,37-39]

In this paper, we present the design and fabrication of a plasmonic nanocavity integrated with $V_{Si}$ defects in 4H-SiC, aimed at enhancing interactions between optical cavity modes and quantum emission. Our design does not require freestanding structures and can be implemented using relatively simple fabrication processes. The fabricated resonator-coupled emitter exhibits a measured Purcell enhancement factor of approximately 48, which slightly exceeds those reported in recent studies. Furthermore, numerical simulations calculate the smallest mode volume of the cavity mode to be about 0.0012 $\mu m^3$, indicating an extremely small, subwavelength mode. Finally, we observed reduced spectral diffusion in the emission well-coupled to the cavity, which is essential for the development of a quantum network based on cavity-integrated color centers.

**Results**

**Design and fabrication of plasmonic cavity integrated SiC spin-active quantum defects.**

**Figure 1A** shows a schematic illustration of a plasmonic nanopan cavity integrated with $V_{Si}$ centers in 4H-SiC. The $V_{Si}$ center can exist as two distinct color centers corresponding to the two inequivalent Si lattice sites in 4H-SiC, with V1 belonging to the cubic site and V2 to the hexagonal site.[12,17] We focus on the V1 variant, which has an optical dipole oriented along the c-axis of the crystal. Moreover, the V1 center is known to have a higher energy orbital transition ZPL (V1′), with an optical dipole orientation along the perpendicular direction to the c-axis, while exhibiting no spin sublevels (**Supporting Information Figure S1**). The cavity-coupled $V_{Si}$ device consists of a SiC disk with an ensemble of $V_{Si}$ centers as optically accessible spin defects, being encapsulated by a silver nanopan. In other words, the bottom and sidewall of SiC disk are covered with silver to construct the nanopan cavity, where a SiC disk-filled silver nanopan structure is expected to support plasmonic cavity modes. We note that the SiC substrate is transparent allowing efficient optical pumping and photoluminescence (PL) collection. Since a smooth metal surface is essential for reducing the scattering loss and increasing the plasmonic propagation length, this inverse type of metal nanodisk (nanopan) structure is expected to be well suited for supporting high-quality (Q) plasmonic modes near the surface of the nanopan.[34] In addition, the optical linewidth of the nanopan cavity is much larger than the spin fine structure of the optical transitions. The coupling between photons and quantum states of emitters is efficient in the strong Purcell regime, in which the emitter interacts primarily with optical modes. In this regime, the spontaneous emission (SE) rate enhancement of the zero-phonon line (ZPL) is given by $F_{ZPL} = F/DW$, where F is the overall Purcell enhancement and DW is the Debye-Waller factor.[40] When the ZPL is coupled to a cavity with quality factor Q and mode volume $V_{mode}$, the enhancement is given by

$$F_{ZPL} = \xi F_{ZPL}^{max} \frac{1}{1+4Q^2(\lambda_{ZPL}/\lambda_{cav}-1)^2}$$

where $F_{ZPL}^{max} = \frac{3}{4\pi^2}\left(\frac{\lambda_{cav}}{n_{eff}}\right)^3 \frac{Q}{V_{mode}}$ is the maximum SE rate enhancement and $\xi = \left(\frac{|\mu \cdot E|}{|\mu||E_{max}|}\right)^2$ which represents the overlap between the optical dipole moment ($\mu$) and the cavity mode electric field (**E**). $F_{ZPL}$ can be enhanced in a cavity with large Q, large $\xi$, and small $V_{mode}$.[29,30] To achieve this, we designed plasmonic cavities using finite-difference time-domain simulations to maximize $F_{ZPL}^{max}$ by optimizing the ratio of $Q/V_{mode}$.

We fabricated the integrated quantum defect/nanopan cavity structure using the following procedure (**Figure 1B** and see **Methods**): First, we form SiC disks with various diameters using electron-beam lithography followed by a dry etching process. Then we remove the hard mask and create $V_{Si}$ centers by implanting He$^+$ ions at a dose of $10^{11}$ /cm$^2$ and an energy of 6 keV. To this end, we deposit a silver layer on the disks.

For measurement, we optically pump through the back side of the transparent SiC substrate. **Figure 1C** shows the highest measured cavity mode showing a Q ~1000, which we estimate by fitting of the fluorescence enhanced by a cavity mode to a Lorentzian lineshape. We note that this cavity is coupled to the phonon sideband (PSB) emission, which is spectrally easier to create a spectral overlap as compared to the narrow ZPL.

**Purcell enhancement of $V_{Si}$ emission.** To assess the optical characteristics of the cavity-coupled $V_{Si}$ device, we optically pump the cavity using a continuous-wave (CW) laser with a wavelength of 780 nm at 10 K (see **Methods**). The pump laser excites $V_{Si}$ centers non-resonantly through their phonon sidebands. We collect the resulting PL using a 0.70 numerical aperture objective lens and send it to avalanche photodiode (APD) detectors or a CCD/spectrometer assembly via optical fiber. Spectral filters were placed in front of the fiber coupler to reject the pump laser.

**Figure 2A** shows the PL spectra measured from on- and off-resonant cavities and bulk SiC substrate. By comparing the emission spectra from different regions, we observe that V1 emission is enhanced by a factor of 23. The measured Q-factor of the cavity resonance near the ZPL of V1 (~861 nm) is approximately 700, showing inhomogeneous broadening of the emission simultaneously. This resonance is readily matched with the narrow ZPL of the emitter. Although V1 emission has an optical dipole orientation along the out-of-plane direction of the substrate, the TM-like resonance mode is well-coupled with this dipole. Furthermore, our cavity is designed to possess many resonant modes with distinct optical properties. To get a better understanding of cavity-enhancement of the fluorescence, we performed PL spectroscopy over cavities with a range of disk diameters between 600 and 1000 nm. We observe strong V1 emission showing peak wavelength of 861.6 nm from a cavity with a diameter of 900 nm (**Figure 2B**). On the other hand, we observe enhanced V1' emission at 859.8 nm from the cavity with a diameter of 650 nm. This emission line is blue shifted by ~4.4 meV relative to the V1 line as illustrated in **Figure 1A**. We note that both emission lines are coupled to different cavity modes since they have different dipole orientations. To investigate the dipole coupling to each emission line as a function of disk diameter, we measure the line intensity of both V1 and V1' emission in 180 individual cavities as a function of disk diameter, ranging from 600 nm to 1000 nm (**Figure 2C**). We obtain a few cavities with wavelengths that are well matched with a target ZPL wavelength, which is helped by the moderate Q-factors of the cavities. There is some variation in the resonant wavelengths of nominally identical cavities because of fabrication variations. When the match is good, the emission intensity is enhanced by the Purcell factor, $F$, which can be obtained by comparing the on- and off-resonant ZPL intensity as previously reported.[29]

$$F = \frac{I_{ZPL,on}}{I_{ZPL,off}} - 1.$$

From this relation, we obtain an estimated maximum enhancement of 12 for the V1' ZPL and 32 for the V1 ZPL, respectively.

Because the emission intensity is sensitive to both the Purcell effect and optical scattering effects such as how efficiently the cavity scatters the ZPL photons into the collection path, we also performed time-resolved PL measurements to obtain the excited-state lifetimes. This method of estimating $F$ purely depends on how strongly the optical dipole is coupled to the cavity. For these measurements, we optically pump the cavity using a picosecond-pulsed Ti:Sapphire laser that is pulse-picked to a repetition rate of 20 MHz and tuned to a wavelength of 780 nm. We record time-resolved fluorescence from cavities that are both on- and off-resonant with $V_{si}$ ZPLs. **Figure 2D** shows a fluorescence lifetime for an on-resonance condition is 2.7±0.24 ns, which is reduced by approximately 3.6 times from the off-resonance condition. This strongly supports that the origin of the fluorescence enhancement is the Purcell effect. For a more quantitative analysis, $F$ can be calculated from lifetime measurements using

$$F = \frac{\tau_0}{\eta}\left(\frac{1}{\tau_{on}} - \frac{1}{\tau_{off}}\right)$$

where $\tau_{on}$ ($\tau_{off}$) refers to the lifetime when the mode is on-resonance (off-resonance) with the ZPL, and $\tau_0$ is the natural lifetime from the emitters in the bulk crystal which is known to be 6.8 ns.[19,29] Here, $\eta$ stands for the branching ratio into the ZPL when the cavity is off-resonance under the same measurement condition that we used. In our measurements, the value of $\eta$ is 3.8% for V1 is used from the measured emission spectrum **(Supporting Information Figure S2)**. Using these values yields an estimated Purcell enhancement of 48, which is consistent with the value estimated from ZPL intensity enhancement.

**Resonance mode analysis.** In-plane polarized resonator modes in nanophotonic cavities (e.g. nanobeam photonic crystal cavities) are predicted to poorly couple to the V1 out-of-plane dipole. Because the spin sublevels that are attractive for quantum information processing correspond to the V1 transition, designing an optical cavity that couples efficiently with an out-of-plane dipole is a key challenge. To analyze the resonance mode of the cavity, we performed numerical eigenvalue studies using finite element methods (FEM) (see **Methods**). We note that the dielectric constant of the silver was modelled by Drude model $\varepsilon(\omega) = \varepsilon_\infty - \omega_p^2/\omega(\omega + i\gamma)$, where $\varepsilon_\infty$ is background dielectric constant and $\omega_p$ is plasma frequency, which is given by 3.1 and 1.4 x $10^{16}$ s$^{-1}$, respectively. The damping collision frequency γ is scaled by a factor of the ratio between the conductivity of a silver at room temperature and at a low temperature.[33,41] **Figure 3A** shows plasmonic whispery gallery modes (WGMs) where the electromagnetic waves propagate along the side wall forming the standing waves. We estimate the theoretical Q-factor of plasmonic WGMs to be approximately 2000 from a nanodisk with a diameter of 900 nm, for which electric field maxima is at the side of the bottom interface between SiC disk and silver cap. The mode volume ($V_{mode}$), which can be defined as $V_{mode} = \frac{\int \varepsilon |E|^2 dV}{max[\varepsilon |E|^2]}$, is calculated as 0.0024 μm$^2$, or 0.45($\lambda/n$)$^3$, which is substantially smaller than previously reported mode volumes for dielectric optical nanocavities. From the simulated mode volume and Q, we find a theoretical maximum Purcell enhancement of 84 in this device. Notably, the electric field vectors of this plasmonic WGMs are calculated to be symmetrically distributed at the side bottom interface between the disk and silver oriented nearly along the z-direction. This orientation of the electric field near the metal, stemming from its fundamental boundary conditions, facilitates efficient modal matching of transverse-magnetic like (TM-like) mode with V1 emission which is oriented along the z-direction (**Figure 3B**). On the other hand, our nanodisk cavity also supports a variety of resonant modes beyond those described.[42] For instance, plasmonic radial modes are excited in the center of the nanodisk near the interface between SiC disk and the silver layer. (**Figure 3C**). The plasmonic radial mode exhibits characteristics similar to the TM-like WGM presented above, in that it interacts well with a dipole oriented in the z-direction (**Figure 3D**). Although the calculated Q of radial mode is lower than that of WGM, it still exhibits a subwavelength mode volume due to the nature of the surface plasmons where the electromagnetic waves are tightly bound to the metal surface. These two types of resonance modes display different resonant wavelengths and spatial distributions of electric field intensity, thereby enhancing the potential for interaction with emitters located at various positions within the cavity. In addition, these plasmonic modes, strongly confined to the interface between the silver layer and SiC, are sensitive to the metal's properties which can lead to a change in Q and resonance wavelength as experimentally shown in **Figure 2D**. Interestingly, there also exist transverse-electric like (TE-like) plasmonic WGMs oriented along the in-plane direction in our metal nanopan cavity (**Supporting Information Figure S3**). If the optical dipole such as V1′, oriented in the in-plane direction, spectrally matches with these modes, we can achieve significant enhancement for this emission as well.

**Spin-dependent optical spectroscopy.** To quantify the optical coherence and address the spin states of the $V_{Si}$ in a plasmonic nanocavity, we perform photoluminescence excitation (PLE) spectroscopy measurements on each emitter while on- and off-resonance with the cavity modes. **Figure 4A** illustrates the energy level structure of the $V_{Si}$ at zero external magnetic field. Ground and excited state manifolds show pairwise degenerate spin levels $m_s = \pm 1/2$ and $m_s = \pm 3/2$, with zero-field splittings of $2 \cdot D_{gs}$ and $2 \cdot D_{es}$, respectively. Each spin-manifold corresponds to spin preserving optical transitions, $A_1$ and $A_2$, respectively. To investigate the excited state structure, we use resonant optical excitation (see **Methods**). A microwave (MW) field at ~4.5 MHz, which is equivalent to the value of $2 \cdot D_{gs}$ of V1 emission, was applied to mix the ground state spin population. In turn, we measure the absorption lines of the emitters by collecting the PSB emissions of the emitters while tuning the wavelength of excitation laser around the 861 nm (ZPL wavelength) across the optically allowed transitions simultaneously. We observe two strong absorption peaks, labelled $A_1$ and $A_2$ from the on-resonant cavity, showing the peak separation of 0.998 GHz, which corresponds to the energy difference between the ground and excited state ZFS (**Figure 4B**). We observed no strong absorption line under the absence of ground state spin mixing.[16] If the electron spin is in a certain state, and is continuously excited along the resonant optical transition, the electron spin will end up in the intersystem crossing, from which the system can decay to all four spin states with an equiprobability. After few cycles, the system cannot be re-excited and is dark.[43] In addition, we observe a similar absorption spectrum from an off-resonant cavity, albeit with ~2 times lower outcoupling fluorescence rates (**Figure 4C**). Interestingly, the Purcell enhancement not only boosts the efficiency of the quantum emission but also improves the quality of the emitted photons by reducing the spectral diffusion.[19] For the on-resonant cavity, the optical transition linewidth is estimated to be 130±3 MHz, corresponding to 87 MHz of spectral diffusion beyond the transform limit, while the off-resonant cavity linewidth is to be 118±4 MHz (corresponds to 102 MHz of spectral diffusion). Spectral diffusion is often caused by excitation of surface-related defects that interact with the emitter. The TM-like cavity mode can be excited with low laser power (<10 nW), resulting in the reduced spectral diffusion. We note that there is a clear reduction in spectral diffusion at 10 K although V1 emission linewidth significantly broadens above cryogenic temperatures owing to relatively small vibronic gap.[15] This reduction is crucial for attaining spectral stability beyond the transform limit, which is important for the development of the quantum network based on cavity-integrated color centers.

**Discussion**

In this study, we have designed and fabricated a plasmonic nanocavity integrated with $V_{Si}$ defects in 4H-SiC for enhanced interactions between optical cavity modes and quantum emitters. Our structure leverages the unique spin-active properties of $V_{Si}$, offering potential advancements for quantum information processing applications. Through careful design and optimization, our cavity exhibits resonant modes that efficiently couple with the distinct optical dipoles of V1 and V1′ centers, enhancing the emission properties and Purcell effect significantly. Finite-difference time-domain simulations were employed to tailor the cavity's optical characteristics, ensuring high-Q plasmonic modes with minimized mode volumes for maximum Purcell enhancement. Experimental validation demonstrated substantial enhancement of V1 emission, indicating effective modal matching and resonance mode coupling. Additionally, our study explores the spin-dependent transitions of $V_{Si}$, highlighting the capability of our plasmonic cavity to facilitate precise control over these states through photoluminescence excitation measurements. The observed reduction in spectral diffusion and enhanced outcoupling rates underscore the potential of our approach for developing robust platforms for quantum technologies. Our findings contribute to the advancement of nanophotonic devices for quantum information, showcasing the importance of integrating spin-active defects with plasmonic cavities for optimized emitter-cavity interactions.

**Methods**

**Device fabrication.** Electron-beam lithography and a subsequent metal mask deposition were used to define the cavity pattern with diameters ranging from 600 to 1000 nm. ICP-RIE (Oxford Cobra) was used to fabricate disks. After removing the metal mask layer using a chemical etchant, He+ ion implantation was performed to generate defects. He+ ions were accelerated at 6 keV and implanted into the sample at a dose of $10^{11}$ cm$^{-2}$. The sample was annealed at 650 °C for 1 hour in a vacuum atmosphere to remove some interstitial-related defects. 500-nm-thick silver layer was deposited on the whole substrates with the SiC disks using electron-beam evaporator (CVC SC4500).

**Optical measurement.** A home-built microscope was used for optical excitation and subsequent fluorescence detection of cavity integrated silicon vacancies. A 780-nm continuous-wave (CW) solid-state laser (Coherent OBIS) was used to optically pump the sample in a He-flow cryostat (Janis) for non-resonant excitation measurements. For fast raster scanning, two piezoelectric steering mirrors were used with a 4-f confocal alignment system. The emission from the sample was collected by a 50× objective lens with a numerical aperture of 0.7 and sent to either an avalanche photodiode (Excelitas SPCM) or a CCD/monochromator (Princeton Instrument Pylon 100BRX). For time-resolved ZPL emission measurements, the sample were optically pumped using a Ti:Sapphire laser (Coherent Chameleon) which is electro-optically modulated by a pulse-picker (ConOptics Model 305). The emission was then sent to the photodiode which was synchronized with a pulse picker, and the decay of the fluorescence was measured by time-correlated single photon counting module (Becker-Hickl SPC130). The Purcell factor was estimated using the measured decay rate enhancement (see the main text). For photoluminescence excitation experiments, we used an external cavity tunable diode laser (Topical DL pro) of which detuning was monitored by Fabry-Perot cavity simultaneously. The collected emission was spectrally filtered in the phonon-sideband region by using tunable filters (Semrock Versa Chrome Edge). Microwaves were generated using signal generators (Stanford Research Systems SG384) and amplified before being applied to the cavities. All the measurements were performed at a cryogenic temperature.

**Numerical simulation.** The eigenmodes of silver nanopan cavities were calculated using the 3D finite-element method (COMSOL Multiphysics). The refractive index of 300-nm-thick dielectric silicon carbide disk on the same substrate with various diameters was set to 2.6. The silver was modelled by Drude model: $\varepsilon(\omega) = \varepsilon_\infty - \omega_p^2/\omega(\omega + i\gamma)$. This model represents the measured dielectric function of silver in the infrared spectrum (800 to 2000 nm), where $\varepsilon_\infty$ represents the background dielectric constant and $\omega_p$ denotes the plasma frequency, assigned values of 3.1 and $1.4 \times 10^{16}$ s$^{-1}$, respectively. The damping collision frequency $\gamma$ was adjusted by a factor corresponding to the ratio of silver's conductivity at room temperature compared to its conductivity at low temperatures to reflect the measurement condition.

**Data availability.** The data that support the findings of this study are available from the corresponding author upon request.

**Acknowledgments**

This work was primarily supported by the National Science Foundation (NSF, OMA-2137828). We also acknowledge support from NSF grant # CMMI-2240267, the Cornell Center for Materials Research (CCMR), an NSF MRSEC (DMR-1719875), the AFOSR (FA9550-23-1-0706) and the ONR (N00014-21-1-2614). This work was completed in part using the shared facilities of the CCMR and Cornell Nanoscale Facility, a NNCI member supported by the NSF (NNCI-2025233). J.-P.S. also acknowledges the support from the NRF grant funded by the Korean government (2022R1A6A3A03064294).

**Author contributions**

J.-P.S. and G.D.F designed the experiments. J.-P.S., J.L., J.C. and B.M. prepared the samples and performed experiments. J.-P.S. performed the simulations. J.-P.S. and G.D.F. analysed the data. J.-P.S. and G.D.F. wrote the paper. All authors discussed the results and commented on the manuscript.

**Competing interests**

The authors declare no competing interests.

**Correspondence and requests for materials** should be addressed to J.-P.S. and G.D.F.

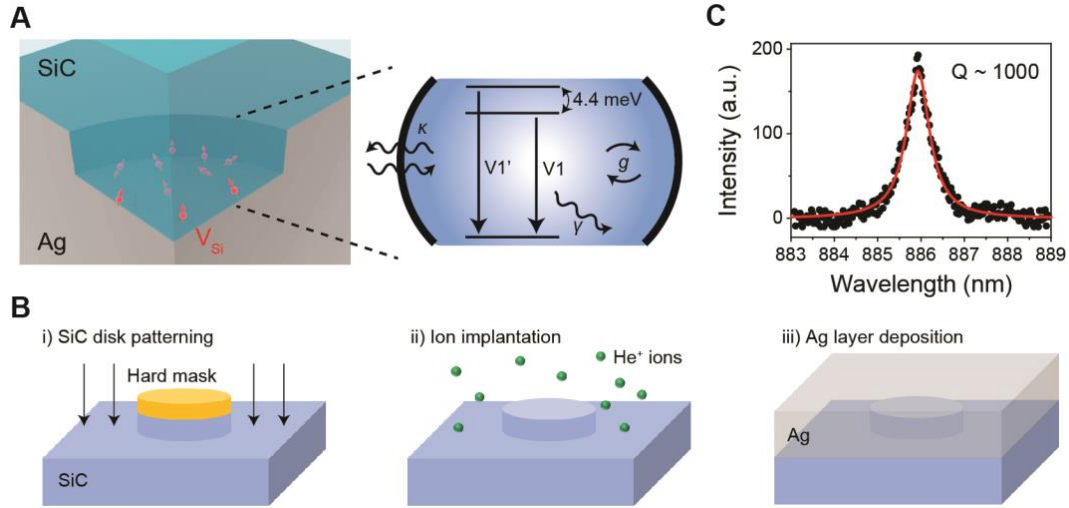

**Fig. 1. Overview of sample design enabling plasmonic cavity integrated V$_{Si}$ centers.** (**A**) Schematic of the SiC nanodisk and the silver nanopan hybrid structure embedding V$_{Si}$ ensembles in the cavity region. The inset shows the V$_{Si}$-nanocavity system showing energy level structure $g$ is cavity-emitter Rabi frequency, $\gamma$ is total spontaneous emission decay rate of the emitter, and $\kappa$ is the cavity decay rate. (**B**) Schematics of the fabrication procedures: (i) SiC disks with a various diameter were formed by using an electron-beam lithography followed by dry etching. (ii) The hard mask was removed, and V$_{Si}$ centers were created by implantation of He$^+$ ions. (iii) The silver layer was deposited on the disks. (**C**) Measured cavity mode coupled to V$_{Si}$ luminescence originated from the PSB showing quality factor of 1,000. A Lorentzian fit is shown in red.

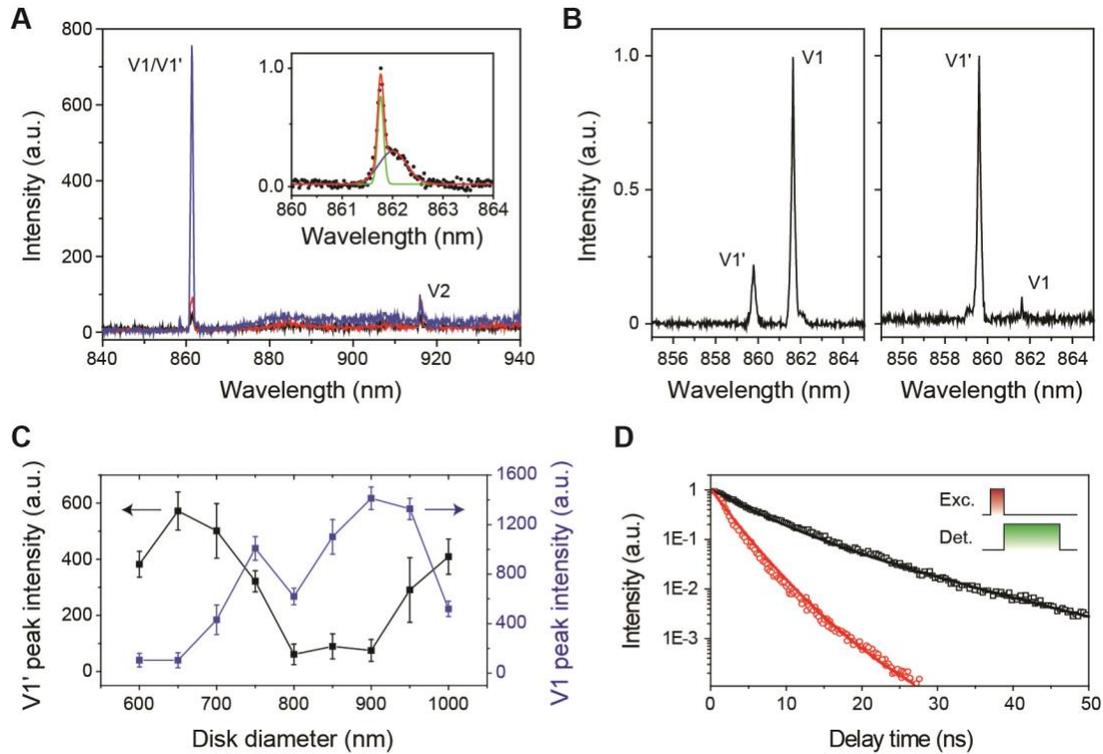

**Fig. 2. Purcell enhancement of the emission from plasmonic cavity integrated V$_{Si}$ centers.** (**A**) A full spectrum of V$_{Si}$ center photoluminescence under non-resonant pumping using a CW 780 nm laser with 135 μW excitation power at T ~ 10 K. The emission is enhanced by a factor of 23 when the cavity is on-resonance with the ZPL (blue) as compared to when the cavity is off-resonance (red). We note that V$_{Si}$ centers are not efficiently excited at this configuration without cavity integration. The inset shows a high-resolution spectrum showing the overlap of an inhomogeneously broadened ZPL and a cavity mode fitted by a double Lorentzian. The linewidth of the fitted cavity mode provides the measured Q of 700 at the wavelength of 862 nm. (**B**) A high-resolution PL spectra showing the enhancement of V1 (left) and V1′ (right), respectively. Each emission is measured from different cavities with a diameter of 900 and 650 nm, respectively. (**C**) A plot of the peak intensity corresponds to V1 (blue) and V1′ (black) from 180 individual cavities

with diameters spaning 600 nm to 1000 nm. (**D**) Excited state lifetime measurements when the cavity is on (red) and off (black) resonance. The data is fitt with a single exponential decay.

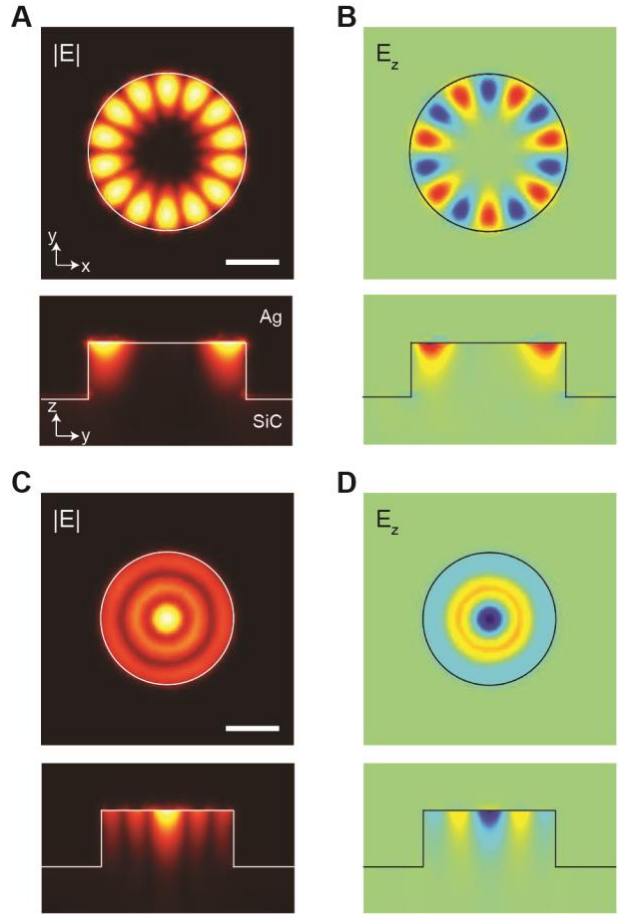

**Fig. 3. Numerical simulation of the cavity resonance modes.** (**A, C**) The calculated normalized electric field intensity profile |E| of the TM-like plasmonic WGM (**A**) and radial mode (**C**), respectively. (**B, D**) The calculated z-components of electric fields profiles from corresponding resonance modes. Each top view of the profile (upper panels) was calculated near the top surface of the SiC disk (10 nm below from the interface). All profiles are showed in the same scale. Scale bar, 300 nm.

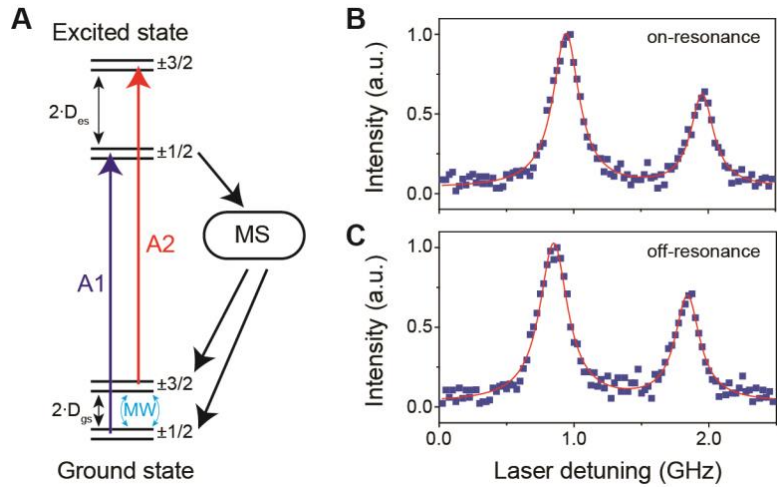

**Fig. 4. Spin-dependent optical absorption of V1 emission ZPL.** (**A**) Energy level diagram of V1 centers. Ground and excited states are spin quartets and the corresponding optical transitions are spin preserving (A1 is for $m_s = 1/2$, A2 is for $m_s = 3/2$, respectively) (**B, C**) Single-scan resonant photoluminescence excitation (PLE) scan from on-resonant cavity (**B**) and off-resonant cavity (**C**), respectively. The fit is based on a double Lorentzian showing the narrower linewidth from A1 transition in both cases. Reduced spectral diffusion was observed from on-resonant cavity.

*Supplementary Information for:*

# Integration of silicon vacancy in silicon carbides with an ultra-small mode-volume plasmonic cavity

Jae-Pil So[1], Jialun Luo[2], Jaehong Choi[1], Brendan McCullian[1], and Gregory D. Fuchs[1*]

[1]School of Applied and Engineering Physics, Cornell University, Ithaca, NY 14850, United States.

[2]Department of Physics, Cornell University, Ithaca, NY 14850, United States.

[*]Corresponding author. E-mail: gdf9@cornell.edu

This PDF file includes:

    Supplementary Figures S1–S3

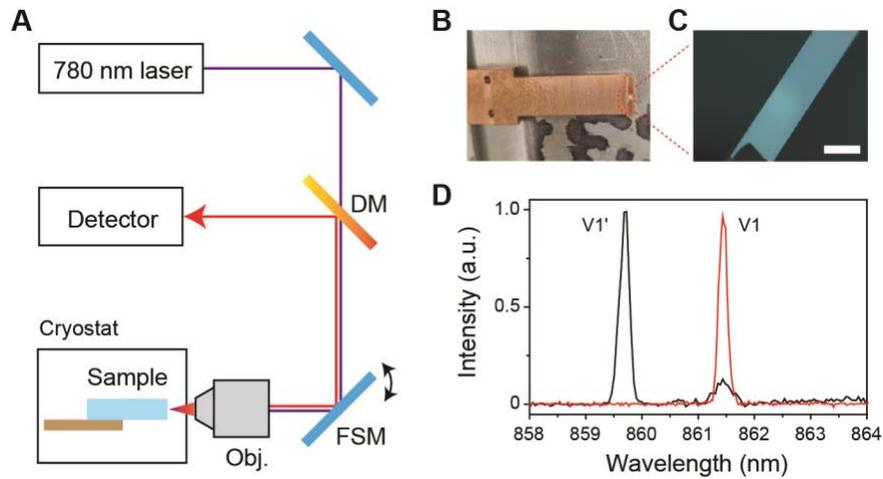

**Fig. S1. The ZPL emission measurement of 4H-SiC substrate mounted on edge.** (**A**) The non-resonant laser with a wavelength of 780 nm passes through a dichroic mirror (DM), and is then sent into the objective (Obj.) by two-axis fast scanning mirrors through a 4f system. The sample is mounted on edge relative to the optical axis in the cryostat. The emitted light travels back through the setup and is sent to an avalanche photodetector or a CCD/spectrometer. (**B**) Real photo of home-made coldfinger extension designed for mounting the sample on edge relative to the optical axis. (**C**) Optical microscope image of the polished edge of 4H-SiC substrate. Scale bar, 100 μm. (**D**) Measured emission spectra from a top mounted sample (black) and an edge mounted one (red), showing clear perpendicular orientations of V1 and V1′ emissions.

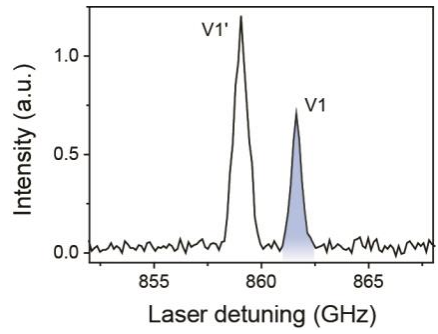

**Fig. S2. Off-resonant zero-photon line (ZPL) emission.** We measured the PL spectrum with the cavity off-resonance and calculated the ratio of the integrated photon counts of the ZPL of the V1 emission (highlighted area) to the total integrated photon counts (including PSB) from the spectrum, yielding a branching ratio $\eta$ of 3.8% into the V1 emission.

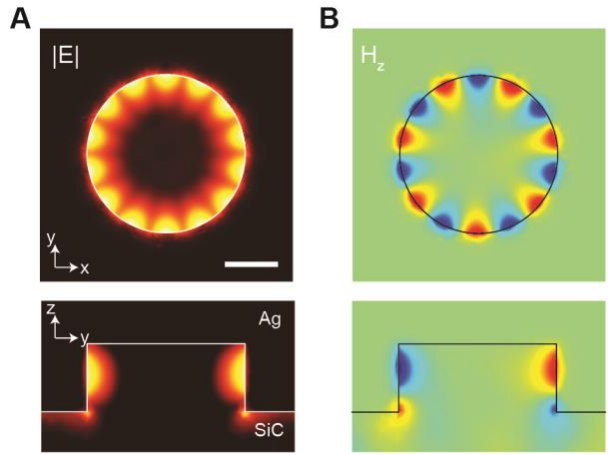

**Fig. S3. Transverse-electric (TE) like whispery gallery mode (WGM).** (**A**) The calculated normalized electric field intensity profile |E| of the TE-like plasmonic WGM. (**B**) The calculated z-components of magnetic fields profiles from corresponding resonance modes, implementing the electric fields oriented along xy-plane. Scale bar, 300 nm.

**Table of Contents Graphic:**

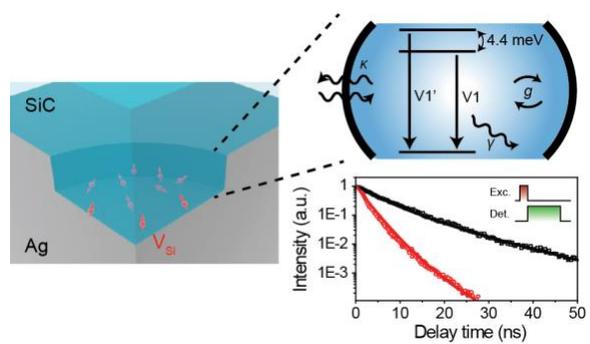